\documentclass[a4paper,10pt]{article}
\usepackage[utf8]{inputenc}
\usepackage[pdftex]{color,graphicx,hyperref}
\usepackage{authblk}

\title{Comparing the hierarchy of author given tags and repository given tags in a large document archive}
\author[1]{Gergely Tibély\thanks{tibelyg@hal.elte.hu}}
\author[2]{Péter Pollner}
\author[2]{Gergely Palla}
\affil[1]{Dept. of Biological Physics, Eötvös University, H-1117 Budapest, Hungary}
\affil[2]{MTA-ELTE Statistical and Biological Physics Research Group, Hungarian Academy of Sciences, H-1117 Budapest, Hungary}

\begin{document}

\maketitle

\begin{abstract}

Folksonomies – large databases arising from collaborative tagging of items by independent users - are becoming an increasingly important way of categorizing information. In these systems users can tag items with free words, resulting in a tripartite item-tag-user network. Although there are no prescribed relations between tags, the way users think about the different categories presumably has some built in hierarchy, in which more special concepts are descendants of some more general categories. Several applications would benefit from the knowledge of this
hierarchy. Here we apply a recent method to check the differences and similarities of hierarchies resulting from tags given by independent individuals and from tags given by a centrally managed repository system. The results from our method showed substantial differences between the lower part of the hierarchies, and in contrast, a relatively high similarity at the top of the hierarchies.

\end{abstract}

Keywords: tag, hierarchy, ontology reconstruction, folksonomy, knowledge mapping

\section{Introduction}

The recent appearance of tags in large online datasets represents a significant innovation in categorisation \cite{mika,spyns,voss}. Tags allow multiple categories for each item, and tagging can be done in a bottom-up approach, in a parallel manner, by several users simultaneously \cite{cattuto,lambiotte,cattuto2}. This feature allows the tagging of huge datasets in a reasonable time. In contrast, traditional hierarchical categorisation typically allows one category per item, and it is done by a few experts, slowing down the process. Also, available categories are restricted in traditional expert-made hierarchies, while user given tags are usually allowed to take any expression deemed relevant by the user.

Although there is no prescribed structure between the tags, it is a reasonable assumption that tags are attached to objects according to hidden hierarchical relations, e.g., ``poodle'' is usually considered as a special case of ``dog''. Consequently, it is an interesting non-trivial task to extract this implicit hierarchy from the co-appearance of tags solely.
Indeed, a number of different methods have already been proposed in the literature, such as aggregation of user-defined shallow hierarchies for obtaining a global hierarchy \cite{shallow,shallow2}, integration of information from as many sources as possible \cite{vandamme}, using a probabilistic criterion to define parent-child relations \cite{schmitz}, applying pairwise similarities to centrality-ordered tags \cite{garciamolina}, or building up the hierarchy from bottom up based on the z-score between the tags \cite{plosonepaper}.

Beside the organisation of different keywords or categories describing a given topic, signs of hierarchy are prevalent in a very wide range of systems. Among others, the transcriptional regulatory network of Escherichia coli \cite{ecoli}, the dominant-subordinate hierarchy among crayfish \cite{crayfish}, the leader-follower network of pigeon flocks \cite{pidgeon}, the rhesus macaque kingdoms \cite{macaque}, neural networks \cite{neuralnets}, technological networks \cite{technets}, social interactions \cite{social1, social2, social3}, urban planning \cite{urban1, urban2}, ecological systems \cite{eco1, eco2}, and evolution \cite{evo1, evo2} all show signs of hierarchical organisation. Different approaches were introduced to uncover hierarchy in networks, including the introduction of hierarchy measures \cite{sneppen, sole1, enys, sole2}, statistical inference of hierarchy \cite{newman} and construction of hierarchical network models \cite{ravasz}.

Here we analyse the hierarchies obtained for the scientific keywords from the Web of Science \cite{wos} by applying a recent generalisation of the method given in Ref.\cite{plosonepaper} presented in \cite{wospaper}. We treat the set of author given tags and the set of repository given tags separately, resulting in two alternative hierarchies. These are compared to each other and also to the 3-level classification of categories given by the Web of Science. The organisation of the paper is the following: in Sect. \ref{s:pre} we introduce the tag hierarchy construction methodology and describe the datasets to which it is applied. The obtained hierarchies are presented in Sect. \ref{s:results},  while the results are discussed in Sect. \ref{s:concl}.

\section{Materials and Methods}  \label{s:pre}

\subsection{Tag hierarchy construction}  \label{s:pre_meth}

In order to obtain a tag hierarchy, we will follow the method described in \cite{plosonepaper} and \cite{wospaper}, for which a quick overview is provided here.

Given a set of objects and each object having a set of tags, the goal is to construct a hierarchy, i.e., a directed acyclic graph (DAG) of the tags, where links are directed from more general concepts to more special ones. Our method constructs a hierarchy in two steps: first the tags are ordered, defining which tag should be placed higher in the hierarchy and which lower, then for each tag an appropriate parent is chosen. Note, that in the second step here we allow to choose more than one parent for a tag, hence the resulting hierarchy can be more complex than a simple tree.

For the reader who is not familiar with the method \cite{plosonepaper} we briefly summarize the main steps below. First we rank first the tags according to the eigenvector centrality of the tag-coappearence graph. Nodes in the co-appearance graph correspond to the tags, and links represent the co-appearances of the tags on the same object. The weights of the links are given by the number of co-appearances. However, when calculating the eigenvector centrality, links having a z-score below a certain threshold value are neglected. The z-score is calculated as the observed number of objects where the two tags co-appear minus the expected number co-occurrences when tags are randomly shuffled. The z-score is normalized by the standard deviation of random co-occurrences,
\begin{equation}
 z_{ij} = \frac{c_{ij} - \mu{ij}}{\sigma_{ij}}
\end{equation}
where $c_{ij}$ is the number of times tags $i$ and $j$ co-appear, $\mu_{ij}$ and $\sigma_{ij}$ are the expected value and standard deviation, respectively, for randomly reshuffled tags.

In the second step the hierarchy is built according to a bottom-up approach, i.e., we look for parents at each tag $i$ in ascending order of their eigenvector centrality. We choose a tag to be the first parent of $i$, when it has higher eigenvector centrality than $i$ and has maximal score among possible parents. The score here is the sum of the z-scores of the links between the candidate parent and the descendants of $i$, and between $i$ itself. Note, that by aggregating the descendants’ z-scores, we take into account much more information than any pairwise similarity metric can provide. Finally, we allow further parents if they have links to $i$ with at least as high z-score as the first parent.

\subsection{Dataset}  \label{s:pre_data}

We study the keywords of scientific papers between 1975 and 2011 obtained from the Web of Science. The dataset contains 35 371 214 papers, which are tagged by three type of tags. The first type (heading) gives a very broad categorisation of the paper, there are only 5 tags of this type: \texttt{Arts \& Humanities}, \texttt{Life Sciences \& Biomedicine}, \texttt{Multidisciplinary Science \& Technology},\\ \texttt{Physical Sciences} and \texttt{Social Sciences}. The second type (category) has 251 more fine-grained scientific areas like \texttt{Chemistry, Analytical} or \texttt{Engineering, Geological}. Tags of the third type are chosen from two sets of specific phrases. One set is composed from the keywords which originated from the authors of the papers. The other set is given by the Web of Science service, and targeted as complementary to the author-given keywords. We will refer to the first keywords as \textit{authorkeywords} and to the other as \textit{woskeywords}. There are a huge number of third-type-tags: the \textit{woskeywords} set contains 2 245 143 phrases and the \textit{authorkeywords} set contains 6 891 089, which are very specific, like \texttt{Zygapophyseal arthritis} or \texttt{H-3 -R-alpha-methylhistamine binding}. Although these keywords are aimed to be complementary on the level of individual papers, still 883 836 of them appear both in the set of \textit{woskeywords} and \textit{authorkeywords}. Finally, we note that the Web of Science does not define any hierarchical relations between the tags, i.e., the ancestors or descendants of the tags are not given in the data set, only the categorization into the three major types is provided.

\section{Results}   \label{s:results}

The aim here is to apply the methodology of Sec. \ref{s:pre_meth} to the data described in Sec. \ref{s:pre_data}, in order to study the differences and similarities of hierarchies resulting from tags given by independent individuals and from tags given by a repository. 
In the first case the input of the hierarchy reconstruction is given by heading, category and \textit{authorkeyword} tags, while in the second case the heading, category and \textit{woskeywords} tags. Note that the general and intermediately general type tags are common in both datasets, and these tags are given by the repository management system. The difference between the independent tagging and centrally managed tagging comes from the most numerous third level tags. We compare below the hierarchies of the two taggings. First we compare the upper most part of the reconstructed DAGs. Then the hierarchy level occupation statistics of the DAGs are compared for each tag types. Finally the horizontal (branching) structures of the DAGs are analysed.

In the reconstructed hierarchies obtained from our method both DAGs had 4 dominant roots at the highest level of the hierarchy, being the ancestors of $99.8\%$-$99.9\%$ of the available tags. The DAGs contain several other non-dominant roots, corresponding to tiny connected components which cover only $0.1\%$-$0.2\%$ of the tags. The four dominant roots coincide with the heading type tags except \texttt{Multidisciplinary Science \& Technology}, which appears as a child of \texttt{Physical Sciences}.

\begin{figure}[!h]
\begin{center}
\includegraphics*[width=\textwidth]{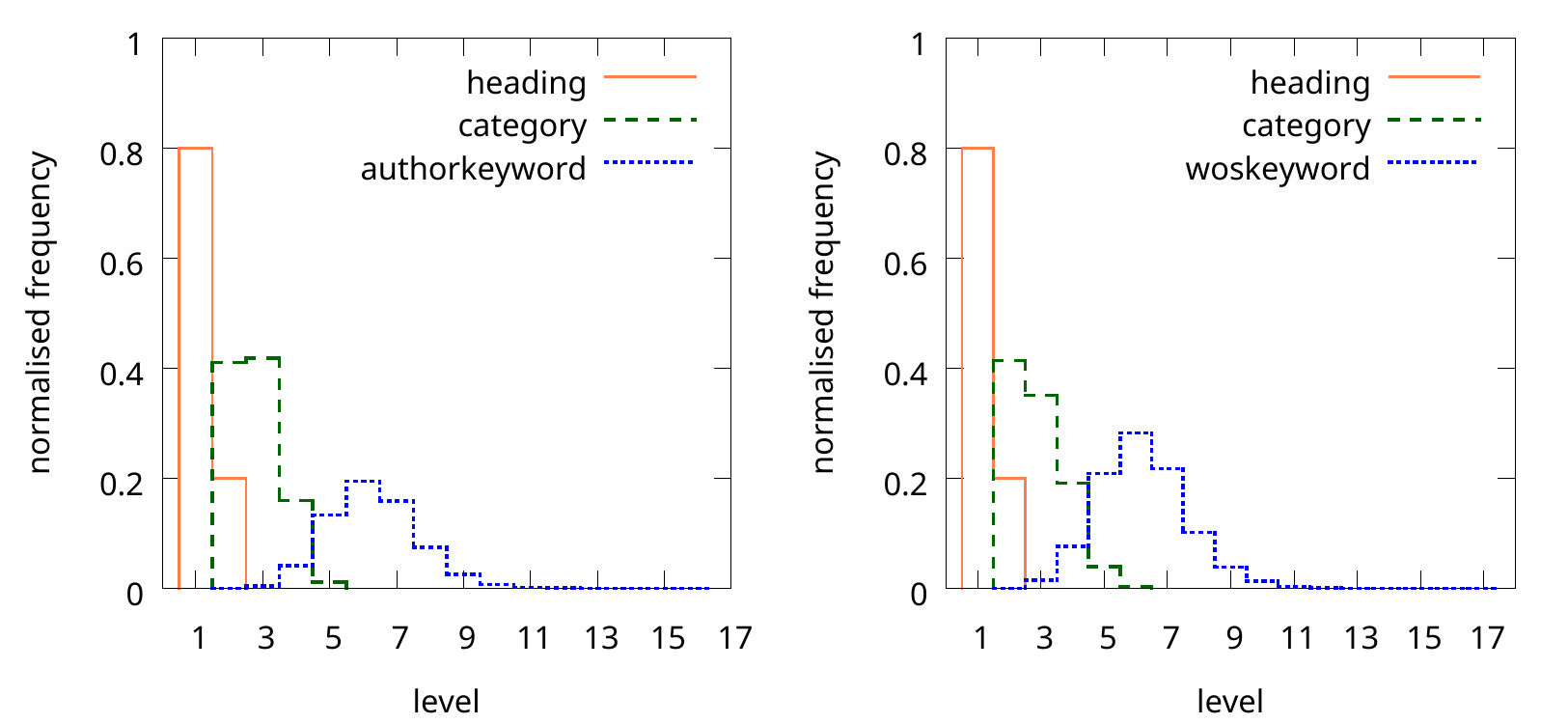}
\end{center}
\caption{Level-wise ratio of tags, for the 3 tag types. Left panel is for the \textit{authorkeyword} DAG, right panel for the \textit{woskeyword} DAG. The distribution is calculated for the tags that are members in at least one of the descendant sets of the 4 dominant roots. Roots are at level 1.}
\label{f:levels}
\end{figure}

Next, we compare the vertical structures of the two DAGs by analysing the hierarchy level distribution of different tag types. A technical difficulty arises from the fact that a tag may belong to more roots, thus it can have more level values depending on the root from which it is counted. Here we classify tags to hierarchy levels according to their closest root, i.e., from the possible level numbers we associate the highest possible level to each tag. The resulting level distributions are shown on Fig. \ref{f:levels}. They indicate that the position in the DAG correlates strongly with the heading-category-(author/wos)keyword classification, i.e., the reconstruction is consistent with the a priori classification of the tags in this respect. However, it is interesting to note, that while tags from different types mostly appear below each other in the expected order, tags from the \emph{same} type also appear below each other -- the reconstruction finds structure within the types. 

The third aspect is the horizontal similarity of the DAGs. Here we analyse whether common members of the DAGs are in similar horizontal position, i.e., having similar descendant subgraphs. Since the DAGs are constructed from the same header and category type tags and the two different keyword tags, we compare the horizontal structure of the two DAGs in two ways:
i) first we restrict the analysis only for those tags, that are common in the two DAGs (header, category and common keywords) ii) secondly we restrict the analysis even more, considering only the header and category type tags, that are common by definition of the DAGs.

\begin{figure}[!ht]
\begin{center}
\includegraphics*[width=\textwidth]{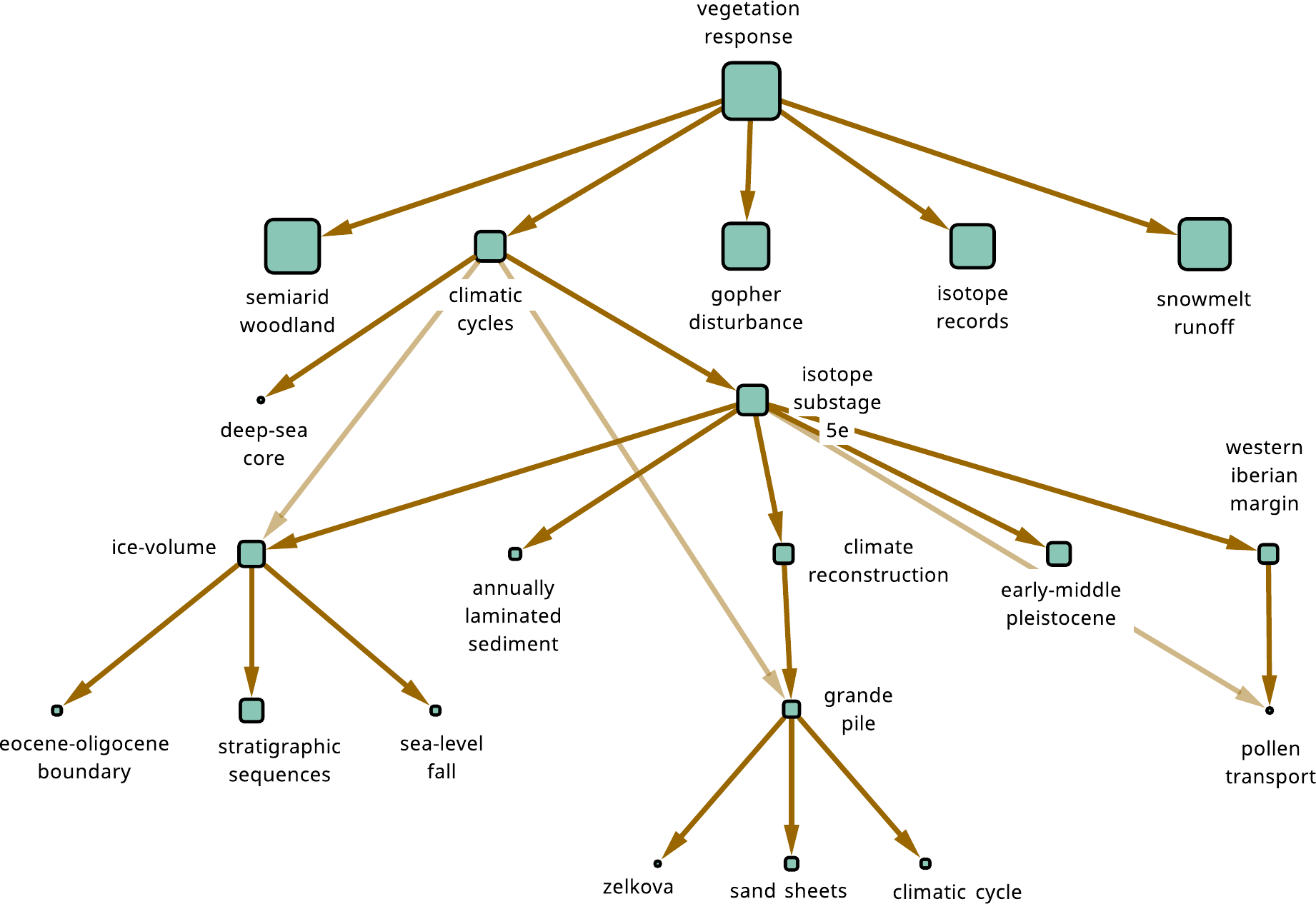}
\vskip0.5cm
\includegraphics*[width=\textwidth]{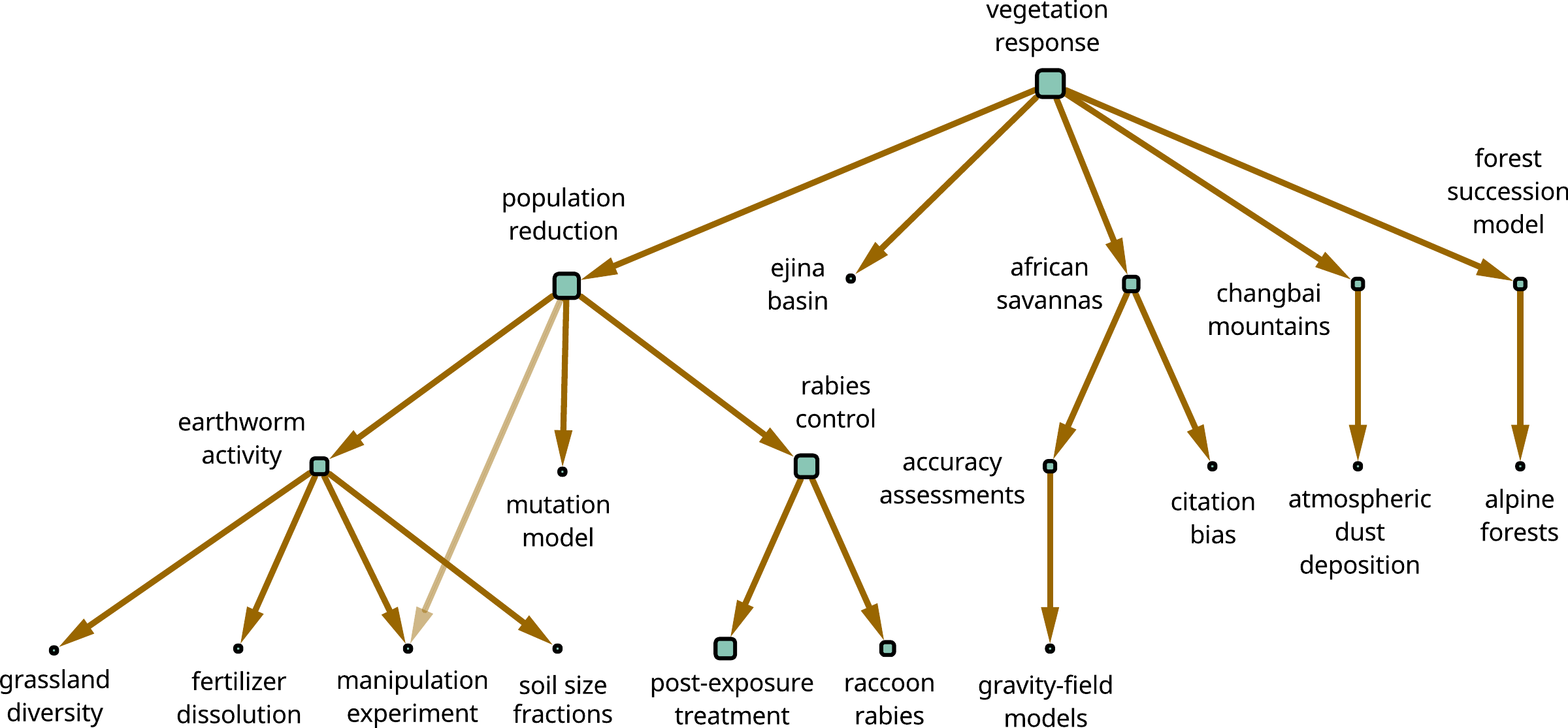}
\end{center}
\caption{Samples from the reduced DAGs (to heading, category and common keywords) with the \textit{woskeywords} (top) and \textit{authorkeywords} (bottom). Node sizes show the number of descendants in the reduced DAGs, on a logarithmic scale.}
\label{f:reduced2common}
\end{figure}

For the first case, where we compare the horizontal position of the common keywords/categories/headers of the two DAGs, we calculated the linearised mutual information-based similarity of \cite{plosonepaper}. The result shows huge dissimilarity with $0.03$ for the mutual information\footnote{The linearised mutual information ranges from 0 to 1.}. A sample of the DAGs is shown on Fig. \ref{f:reduced2common}, around ``vegetation response''. In both DAGs, related tags appear below the chosen tag, however, according to Fig. \ref{f:reduced2common}, descendants in one DAG differ from descendants in the other. These results are in accordance with the complementary nature of the \textit{authorkeywords} and \textit{woskeywords}.

\begin{figure}[!ht]
\begin{center}
\includegraphics*[width=\textwidth]{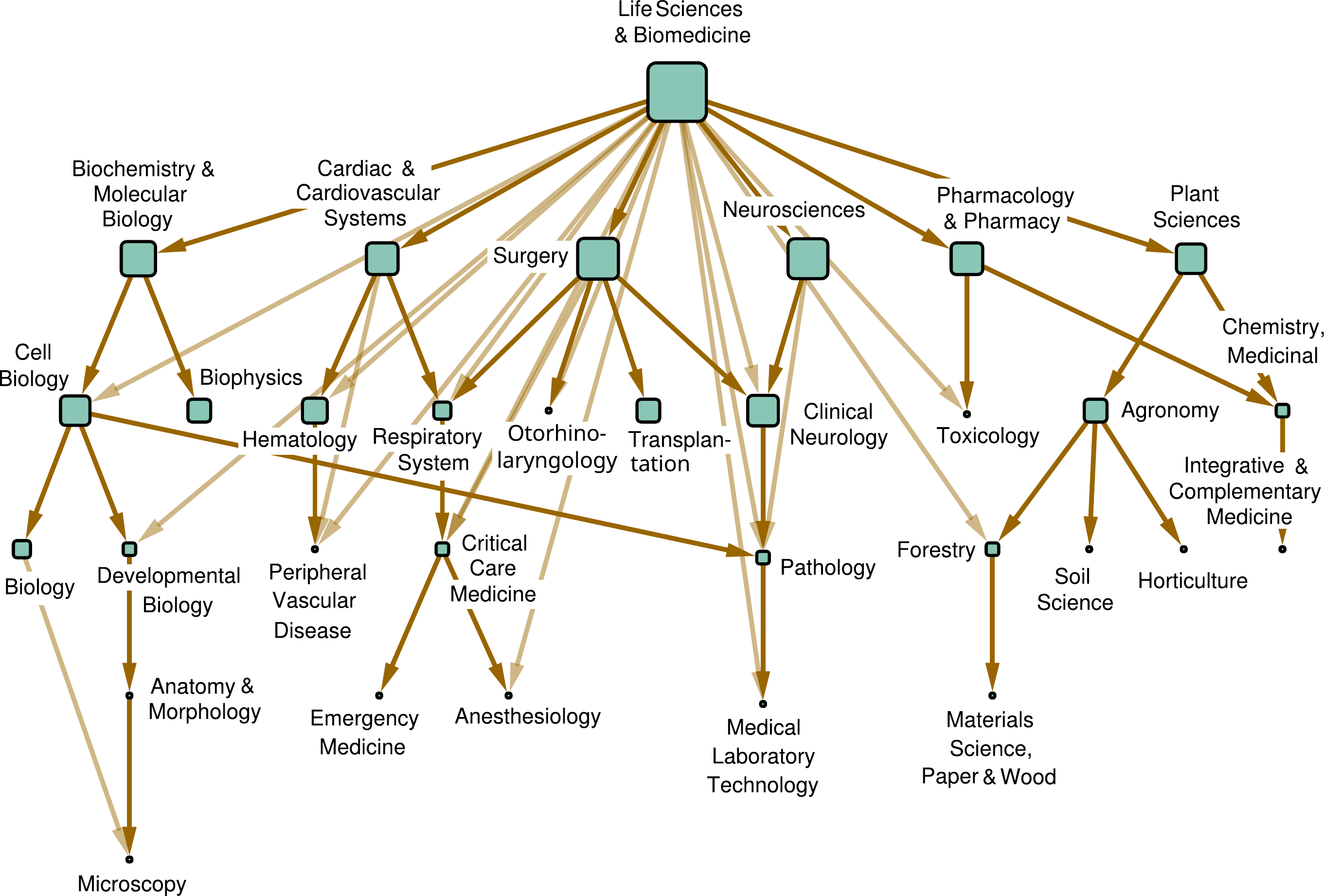}
\includegraphics*[width=0.85\textwidth]{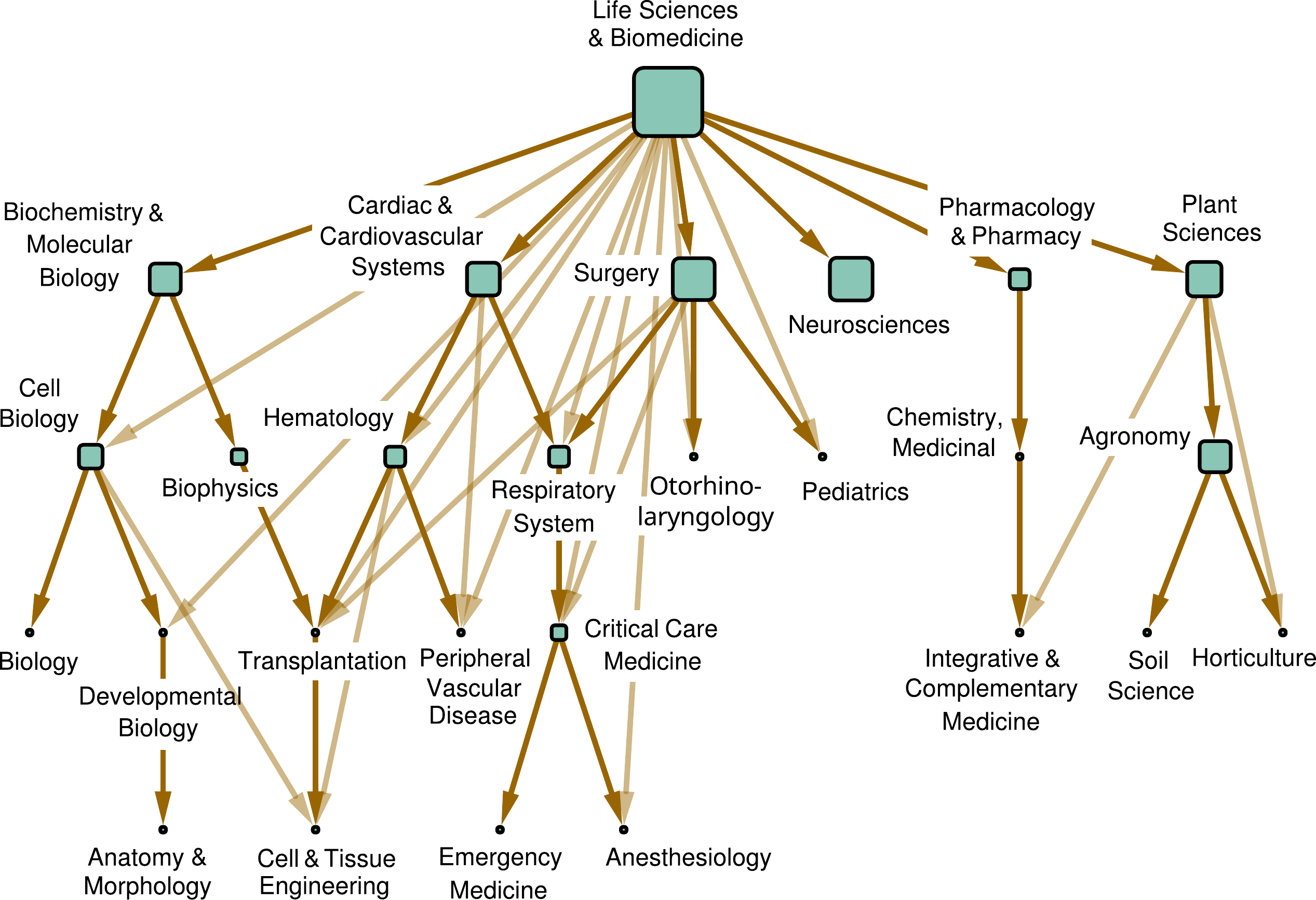}
\end{center}
\caption{Samples from the reduced DAGs (to heading and category) of the \textit{woskeywords} (top) and \textit{authorkeywords} (bottom) based reconstructions. Reduction left only the 256 heading and category tags. Node sizes show the number of descendants in the reduced DAGs, on a logarithmic scale.}
\label{f:reduced2hc}
\end{figure}

If we restrict the calculation of the mutual information to the header and category tags only, the similarity jumps to $0.89$, showing that the relations between general tags are quite robust, indeed, the hierarchies are built bottom-up, where the bottom parts are very different. Samples of these reduced DAGs are visualised on Fig. \ref{f:reduced2hc}. They display a few branches below \texttt{Life Sciences \& Biomedicine}, like \texttt{Biochemistry \& Molecular Biology}, \texttt{Cardiac \& Cardiovascular Systems} or \texttt{Plant} \texttt{Sciences}. The two sub-figures show that \texttt{Neurosciences, Plant Sciences, Biophysics} and \texttt{Agronomy} have more children in the \textit{woskeyword} DAG, while \texttt{Hematology} is also connected to \texttt{Transplantation} in the \textit{authorkeyword} DAG.\\
Note that the reconstruction strongly depends on the descendants of each tag, especially for tags having several descendants, thus the difference between the \textit{authorkeywords} and \textit{woskeywords} could have led to very different structure at the top of the DAG \cite{plosonepaper}. The very high similarity at the top of the hierarchy compared to the low similarity for the first case indicates, that the differences between the \textit{authorkeywords} and the \textit{woskeywords} result differences on the low levels of the hierarchy, while this difference does not propagate to the highest levels.

\section{Discussion}  \label{s:concl}

Tag-based categorisation of large online datasets is becoming increasingly wide\-spread. They allow free word tagging, multiple categories for items and user-based processing in a parallel manner instead of centralised expert-based processing. Although the tags have no predefined relations, it is reasonable to assume that users think to an extent in hierarchical relations between tags, i.e., using some tags as special cases of other, more general tags.

Here we applied a recently introduced hierarchy construction method to keywords of scientific papers from the Web of Science. Tags were pre-organised by the Web of Science into 3 types, from the very general to the very special. For the most special type, 2 different sets of keywords were obtained, author-given and repository-given. Accordingly, two different hierarchies were constructed, each time using one of these sets as the special type, accompanied by the more general tags.

First, the structures of the obtained hierarchies were compared to the 3 predefined tag types. Good correspondence was found here. For the most general type, 4 out of the 5 member tags appeared as level 1 roots in the constructed hierarchies (\texttt{Arts \& Humanities}, \texttt{Life Sciences \& Biomedicine}, \texttt{Physical Sciences} and \texttt{Social Sciences}), the fifth one being an immediate child of one of them (\texttt{Multidisciplinary Science \& Technology}). The intermediate type tags populated the next levels in the hierarchies, and members of the most specific type were at the lowest levels. An interesting observation is that the tags were organised to significantly more levels than three, indicating that there is structure within the predefined types.

Second, the two constructed hierarchies, using two different set of special keywords, were compared to each other. The hierarchies were reduced to the tags common in both of them, in order to make direct comparison possible. It was found that the organisation of the tags are very different, their similarity scoring $0.03$ on a [0,1] scale. This is in accordance with their purpose, i.e., for each individual paper \textit{woskeywords} are aimed to be complementary to the \textit{authorkeywords} \cite{isi}. On the other hand, when reducing the hierarchies only to the general and intermediately general type tags, a much higher $0.89$ similarity was obtained, in spite of the fact that the hierarchies were constructed bottom up, allowing different lower levels resulting in different high levels. Interestingly, while the lower parts of the hierarchies were different, the more general tags were organised in a significantly similar way.

\section{Acknowledgments}

The research was partially supported by the European Union and the European Social Fund through project FuturICT.hu (grant no.:TAMOP-4.2.2.C-11/1/KONV-2012-0013) and by the Hungarian National Science Fund (OTKA K105447). The funders had no role in study design, data collection and analysis, decision to publish, or preparation of the manuscript. The authors declare no conflict of interest.


\begin{thebibliography}{00}

\bibitem{mika} Mika P (2005) Ontologies are us: A unified model of social networks and semantics. In: \textit{In International Semantic Web Conference} 3729522–536.

\bibitem{spyns} Spyns P, Moor AD, Vandenbussche J, Meersman R (2006) From Folksologies to Ontologies: How the Twain Meet. In: \textit{In Proceedings of OTM Conferences} 1738–755.

\bibitem{voss} Voss J (2007) Tagging, folksonomy \& Co - renaissance of manual indexing? \texttt{ArXiv:cs/0701072v2}.

\bibitem{cattuto} Cattuto C, Loreto V, Pietronero L (2007) Semiotic dynamics and collaborative tagging. \textit{Proc Natl Acad Sci USA} \textbf{104}: 1461–1464.

\bibitem{lambiotte} Lambiotte R, Ausloos M (2006) Collaborative tagging as a tripartite network. \textit{Lect Notes in Computer Sci} \textbf{3993}: 1114–1117.

\bibitem{cattuto2} Cattuto C, Barrat A, Baldassarri A, Schehr G, Loreto V (2009) Collective dynamics of social annotation. \textit{Proc Natl Acad Sci USA} \textbf{106}: 10511–10515.

\bibitem{shallow} Plangprasopchok A, Lerman K (2009) Constructing folksonomies from user-specified relations on flickr. In: \textit{Proceedings of the World Wide Web conference} pp. 781–790.

\bibitem{shallow2} Plangprasopchok A, Lerman K, Getoor L (2011) A probabilistic approach for learning folksonomies from structured data. In: \textit{Fourth ACM International Conference on Web Search and Data Mining (WSDM)} pp. 555–564.

\bibitem{vandamme} Damme CV, Hepp M, Siorpaes K (2007) Folksontology: An integrated approach for turning folksonomies into ontologies. \textit{Social Networks} \textbf{2}: 57–70.

\bibitem{schmitz} Schmitz P (2006) Inducing ontology from flickr tags. In: \textit{Proc. of Collaborative Web Tagging Workshop at the 15th Int. Conf. on World Wide Web (WWW)}.

\bibitem{garciamolina} Heymann P, Garcia-Molina H (2006) Collaborative creation of communal hierarchical taxonomies in social tagging systems. Technical report, Stanford InfoLab. URL \texttt{http://ilpubs.stanford.edu:8090/775/}.

\bibitem{plosonepaper} Tibély G, Pollner P, Vicsek T, Palla G (2013) Extracting Tag Hierarchies. \textit{PLoS ONE} \textbf{8}, e84133.

\bibitem{ecoli} H. W. Ma, J. Buer and A. P. Zeng (2004) Hierarchical sructure and modules in the Escherichia coli transcriptional regulatory network revealed by a new top-down approach. \textit{BMC Bioinformatics} \textbf{5}:199.

\bibitem{crayfish} C. Goessmann, C. Hemelrijk and R. Huber (2000) The formation and maintenance of crayfish hierarchies: behavioral and self-structuring properties. \textit{Behavioral Ecology and Sociobiology} \textbf{48}:418-428.

\bibitem{pidgeon} M. Nagy, Z. Ákos, D. Biro and T. Vicsek (2010) Hierarchical group dynamics in pigeon flocks. \textit{Nature} \textbf{464}:890-893.

\bibitem{macaque} H. Fushing, M. P. McAssey, B. Beisner and B. McCowan (2011) Ranking network of captive rhesus macaque society: A sophisticated corporative kingdom. \textit{PLoS ONE} \textbf{6}:e17817.

\bibitem{neuralnets} M. Kaiser, C. C. Hilgetag and R. Kötter (2010) Hierarchy and dynamics of neural networks. \textit{Front. Neuroinform.} \textbf{4}:112.

\bibitem{technets} D. Pumain (2006) Hierarchy in Natural and Social Sciences. \textit{Methodos Series 3} (Springer Netherlands, Dodrecht, The Netherlands).

\bibitem{social1} R. Guimer\`a, L. Danon, A. Díaz-Guilera and F. Giralt and A. Arenas (2003) Self-similar community structure in a network of human interactions. \textit{Phys. Rev. E} \textbf{68}: 065103.

\bibitem{social2}P. Pollner, G. Palla and T. Vicsek (2006) Preferential attachment of communities: The same principle, but a higher level. \textit{Europhys. Lett.} \textbf{73}: 478–484.

\bibitem{social3} S. Valverde and R. V. Solé (2007) Self-organization versus hierarchy in open-source social networks. \textit{Phys. Rev. E} \textbf{76}:046118.

\bibitem{urban1} P. R. Krugman (1996) Confronting the mystery of urban hierarchy. \textit{J. Jpn. Int. Econ.} \textbf{10}: 399-418.

\bibitem{urban2} M. Batty and P. Longley (1994) Fractal Cities: A Geometry of Form and Function (Academic, San Diego).

\bibitem{eco1} H. Hirata and R. Ulanowicz (1985) Information theoretical analysis of the aggregation and hierarchical structure of ecological networks. \textit{J. Theor. Biol.} \textbf{116}:321-341.

\bibitem{eco2} J. Wickens and R. Ulanowicz (1988) On quantifying hierarchical connections in ecology. \textit{J. Soc. Biol. Struct.} \textbf{11}:369-378.

\bibitem{evo1} N. Eldredge (1985) Unfinished Synthesis: Biological Hierarchies and Modern Evolutionary Thought. (Oxford Univ. Press, New York).

\bibitem{evo2} D. W. McShea (2001) The hierarchical structure of organisms. \textit{Paleobiology} \textbf{27}:405-423.

\bibitem{sneppen} A. Trusina, S. Maslov, P. Minnhagen, K. Sneppen (2004) Hierarchy measures in complex networks. \textit{Phys. Rev. Lett.} \textbf{92}:178702.

\bibitem{sole1} B. Corominas-Murtra, C. Rodríguez-Caso, J. Go\~ni and R. Solé (2011) Measuring the hierarchy of feedforward networks. \textit{Chaos} \textbf{21}:016108.

\bibitem{enys} E. Mones, L. Vicsek and T. Vicsek (2012) Hierarchy Measure for Complex Networks. \textit{PLoS ONE} \textbf{7}:e33799.

\bibitem{sole2} B. Corominas-Murtra, J. Go\~ni, R. V. Solé and C. Rodríguez-Caso (2013) On the origins of hierarchy in complex networks. \textit{Proc. Natl. Acad. Sci. USA} \textbf{110}:13316-13321.

\bibitem{newman} A. Clauset, C. Moore and M. E. J. Newman (2008) Hierarchical structure and the prediction of missing links in networks. \textit{Nature} \textbf{453}:98–101.

\bibitem{ravasz} E. Ravasz, A. L. Somera, D. A. Mongru, Z. N. Oltvai, A.-L. Barabási (2002) Hierarchical Organization of Modularity in Metabolic Networks. \textit{Science} \textbf{297}:1551–1555.

\bibitem{wos} Isi web of knowledge. \texttt{http://scientific.thomson.com/isi/} (Date of access: 01/01/2012).

\bibitem{wospaper} Palla G, Tibély G, Mones E, Pollner P, Vicsek T (2015) Hierarchical networks of scientific journals. \texttt{arXiv:1506.05661} To appear in Palgrave Communications.

\bibitem{isi} \texttt{http://interest.science.thomsonreuters.com/content/WOKUserTips-}\texttt{201010-SEA} (Date of access: 15/12/2014).


\end{thebibliography}
\end{document}